\documentclass[aps,prx,reprint,10pt]{revtex4-2}

\usepackage{graphicx,amsmath,amssymb,dcolumn,physics,color,bm}

\newcommand{\kB}{k_\mathrm{B}}

\newcommand{\physics}{Department of Physics and Quantitative Biology Institute, Yale University, New Haven, CT, USA}

\begin{document}

\title{Nonequilibrium fluctuations could explain the hot mitochondria paradox}

\author{Matthew P. Leighton}
\email{matthew.leighton@yale.edu}
\author{Benjamin B. Machta}
\affiliation{\physics}

\date{\today}

\begin{abstract}
Several different experiments have found apparent temperature differences of up to 10 K between the interior of mitochondria and the rest of the cellular environment. The physics of heat conduction, however, precludes the steady-state existence of a temperature difference greater than the microkelvin scale within the cell, absent a far greater energy budget. Nearly a decade after the first reported measurement, this apparent paradox remains unexplained. Here, we propose an explanation which allows the elevated temperature measurements to coexist with the theoretical impossibility of a true 10 K steady-state temperature difference. We suggest that nonequilibrium fluctuations within the cellular environment, combined with molecular temperature probes sensitive to specific ranges of frequencies, can lead to apparent 10 K temperature differences with a biologically attainable power budget.
\end{abstract}

\maketitle

How hot are mitochondria? This seemingly innocuous question has led to much controversy over the last decade. In 2018, it was reported that experimental measurements of mitochondrial temperature using a fluorescent molecular probe found that the interior of the mitochondria was maintained at a temperature $\gtrsim10$ K hotter than the rest of the cellular environment~\cite{chretien2018mitochondria}. Subsequent experiments using a variety of different measurement techniques have similarly found mitochondrial temperatures elevated by $1$--$10$ K with respect to the rest of the cell, either at steady state, or transiently following chemically-induced proton gradient dissipation~\cite{savchuk2019gfp,romshin2022heat,di2022spatiotemporally}. Ref.~\cite{macherel2021conundrum} provides an excellent overview of relevant experimental findings.

One possibility is that the solvent within the mitochondrial matrix could truly be $1$--$10$ K hotter than the rest of the cell. A steady-state temperature difference of this magnitude across a nanometer-scale membrane, however, is incompatible with the constraints of biology~\cite{baffou2014critique}. Heating a mitochondria-sized sphere in water by $10$ K would require $\sim\!10^7$ pW. The energy budgets of a typical cell ($\sim\!100$ pW) or mitochondrion ($\sim\!0.1$ pW)~\cite{milo2010bionumbers}, by contrast, limit such temperature differences to the realm of microkelvin~\cite{macherel2021conundrum}. Even conjecturing the existence of active heat pumps is insufficient by far to explain the observed temperature difference~\cite{tkachenko2026hot}.

The second class of proposed explanations consider the unique environment of the mitochondria, which exhibits extreme pH levels, a high degree of molecular crowding, and confined geometry, arguing that perhaps the temperature probes' calibrations fail in such an environment. Other attempts to resolve the paradox have postulated large local transient temperature spikes at steady state, still requiring a real increase in solvent temperature inside the mitochondria~\cite{fahimi2022hot}. Ultimately, the conundrum of the hot mitochondria remains unresolved~\cite{matta2026revisiting}.

Here, we propose an alternative explanation which allows the experimental temperature measurements to coexist with the theoretical impossibility of a steady-state temperature difference on the order of $10$ K. The cellular environment is far from equilibrium, with fluctuations violating the fluctuation-dissipation theorem. Within such an environment, a temperature probe's coordinates will be driven out of equilibrium in a manner dependent on which frequencies of fluctuations they are most sensitive to~\cite{cugliandolo1997energy,fodor2015activity}. Thus two different temperature probes, both properly calibrated in equilibrium environments, can nonetheless report different temperature measurements. We explore a toy model of such a setup, and show that the reported mitochondrial temperature elevation can be achieved through nonequilibrium fluctuations with a biologically plausible power budget.

\subsection*{Thermometry far from Equilibrium}
At equilibrium, the quantity called ``temperature" is both well-understood and easy to define. One of the triumphs of classical thermodynamics and statistical mechanics is that many different definitions of temperature all describe precisely the same quantity at equilibrium~\cite{callen2006thermodynamics}.

In classical thermodynamics, temperature is defined through either the zeroth law, or the derivative of entropy with respect to energy. In statistical mechanics, temperature is instead defined using the Boltzmann distribution or equipartition theorem. While these definitions all involve static quantities, temperature can also be defined in terms of dynamics through the fluctuation-dissipation theorem (FDT). The FDT relates the power spectrum $S_x(\omega)$ of a coordinate $x(t)$ at frequency $\omega$, the Fourier transform $\chi_x(\omega)$ of that coordinate's response function to perturbations of frequency $\omega$, and the equilibrium temperature $T$, as
\begin{equation}\label{eq:FDT}
S_x(\omega) = -\frac{2\kB T}{\omega}\mathrm{Im}\,\chi_x(\omega).
\end{equation}
If both $S_x(\omega)$ and $\chi_x(\omega)$ are measured for a given environment and probe coordinate $x$, \eqref{eq:FDT} can then be used to define an effective temperature $T_\mathrm{eff}(\omega)$~\cite{cugliandolo1997energy,fodor2015activity,dieterich2015single}.

The interior of the cell is far from equilibrium, lying outside the realm of equilibrium statistical physics. Of greatest relevance to the present discussion, the fluctuations inside a cell do not generally satisfy the FDT, and can exhibit highly inhomogeneous power spectra~\cite{mizuno2007nonequilibrium,gallet2009power}. Measurements in different contexts have shown that cellular fluctuations can lead to large effective temperatures at certain frequencies~\cite{weber2012nonthermal,narinder2026time}.

In addition to power-spectrum inhomogeneities, which depend both on the environment and the response function of the specific temperature probe, the measured observable will be most sensitive to certain frequencies of fluctuations. We postulate that the measured temperature can be written as an average of the frequency-dependent effective temperature over a probe kernel which we denote $g(\omega)$:
\begin{equation}\label{eq:Tmeasdef}
T_\mathrm{meas} = \int_{0}^\infty \frac{\mathrm{d} \omega}{\pi} g(\omega)T_\mathrm{eff}(\omega).
\end{equation}
A temperature probe accurately calibrated at equilibrium will have a kernel $g(\omega)$ that integrates to one. Thus a temperature probe properly calibrated at equilibrium need not return the true temperature $T$ in an environment with nonequilibrium fluctuations, and two probes with different kernels need not report the same measured temperature in the same environment. In the following section we illustrate this phenomenon explicitly with a simple model. 

\subsection*{A Simple Model}
\begin{figure}[t]
    \centering
    \includegraphics[width=\linewidth]{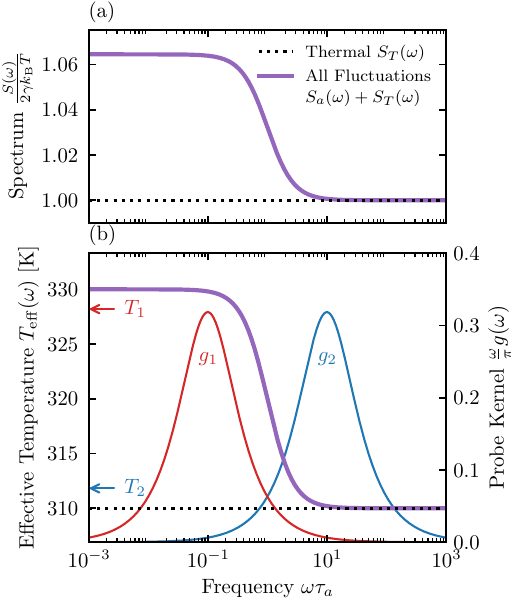}
    \caption{Fluctuation spectrum and effective temperature. (a) Power spectrum for thermal noise (dotted line) combined with Ornstein-Uhlenbeck noise (purple). (b) Frequency-dependent effective temperature (purple), and kernels (scaled by $\omega/\pi$ to show log-weighting) for probes with two different relaxation times (red: $\tau_p = 10\tau_a$, blue: $\tau_p = \tau_a/10$). Red and blue arrows indicate the temperatures measured by the two probes. We take $T = 310$ K and $D_a/\kB \gamma = 20$ K.
    \vspace{-3mm}
    }
    \label{fig:fig1}
\end{figure}

As a simple model for a temperature probe, consider a single overdamped coordinate $x(t)$ with friction coefficient $\gamma$, in a quadratic potential energy landscape with stiffness $\kappa$. The probe experiences both equilibrium fluctuations from a thermal bath at temperature $T$, and additional nonequilibrium fluctuations. The dynamics are
\begin{equation}
\gamma\dot{x} = -\kappa x + \xi_T(t) + f_a(t).
\end{equation}
Here $\xi_T$ is thermal white noise with spectrum $S_T(\omega) = 2\gamma \kB T$. We model the nonequilibrium fluctuations $f_a(t)$ as Ornstein-Uhlenbeck noise with strength $D_a$ and correlation time $\tau_a$. They thus have zero mean ($\langle f_a(t)\rangle=0$), but non-constant power spectrum $S_a(\omega) = 2D_a/(1 + \tau_a^2\omega^2)$ [Fig.~\ref{fig:fig1}(a)].

The probe has response function $\chi_x(\omega) = 1/(\kappa + i\gamma\omega)$, and the power spectrum of the $x$ coordinate is $S_x(\omega)=|\chi_x(\omega)|^2[S_T(\omega) + S_a(\omega)]$. The effective temperature is thus
\begin{equation}
T_\mathrm{eff}(\omega) = T + \frac{D_a/(\kB \gamma)}{1 + \tau_a^2\omega^2}.
\end{equation}

For this particular model, the steady-state distribution of the probe coordinate $x$ is always Gaussian, and thus indistinguishable from an equilibrium Boltzmann distribution at some finite temperature. Suppose then that we take as an observable any quantity dependent only on the steady-state distribution, and properly calibrate it across a range of equilibrium environments. Inserting the probe into an environment with nonequilibrium fluctuations as depicted in Fig.~\ref{fig:fig1}(a), one would find upon measuring the observable that it is entirely consistent with an equilibrium temperature:
\begin{subequations}
\begin{align}
T_\mathrm{meas} & = \int_{0}^\infty \frac{\mathrm{d}\omega}{\pi} \underbrace{\frac{2\tau_p}{1 + \omega^2\tau_p^2}}_{g(\omega)}T_\mathrm{eff}(\omega)\\
& = T + \frac{\tau_p}{\tau_p+\tau_a}\frac{D_a}{\gamma \kB }.
\end{align}
\end{subequations}
Here we have identified the probe kernel $g(\omega)$, which depends only on its relaxation time $\tau_p = \gamma/\kappa$. For $\tau_p\ll\tau_a$, we have $T_\mathrm{meas}\approx T$, while for larger $\tau_p$ the measured temperature increases up to $T_\mathrm{meas}= T + D_a/\kB \gamma$ as $\tau_p\to\infty$. Thus two probes with different relaxation times can potentially measure very different temperatures [Fig.~\ref{fig:fig1}(b)].

\begin{figure}[t]
    \centering
    \includegraphics[width=\linewidth]{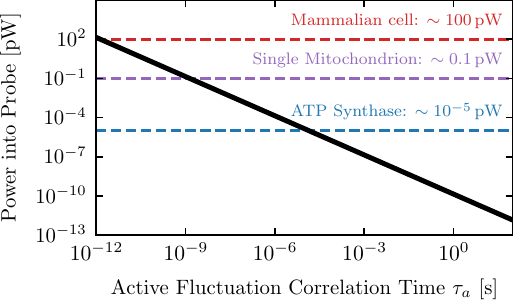}
    \caption{Probe power dissipation required to maintain an apparent $10$ K temperature difference (black), from \eqref{eq:powerbudget}. Horizontal dashed lines indicate order-of-magnitude estimates for the power budgets of an entire mammalian cell (red), a single mitochondrion (purple), and a single ATP synthase (blue), based on Ref.~\cite{milo2010bionumbers}. 
    \vspace{-3mm}
    }
    \label{fig:fig2}
\end{figure}

\newpage
The remaining question is whether this effective temperature can be achieved with a power budget in the realm of possibility. To answer it, we compute the power from the nonequilibrium fluctuations into the probe, which at steady state is then dissipated as heat into the thermal environment,
\begin{equation}\label{eq:powerbudget}
P_a \equiv \left\langle f_a(t)\circ\dot{x}(t)\right\rangle = \kB \frac{T_\mathrm{meas} - T}{\tau_a}.
\end{equation}
Here ``$\circ$" denotes Stratonovich multiplication. 
Note that $P_a$ is the power dissipated through the probe, not the total cost of maintaining the nonequilibrium fluctuations.

For our simple model, the power into the probe depends only on the temperature difference between the measured and equilibrium temperatures, and the correlation time of the nonequilibrium fluctuations. For a fast correlation time on the order of picoseconds, a temperature difference  $T_\mathrm{meas}-T= 10$~K would require approximately the entire power budget of a typical mammalian cell. For longer correlation times on the order of tens to hundreds of microseconds, however, the effect can be powered by less than the output of a single F$_\mathrm{o}$F$_1$-ATP synthase machine [Fig.~\ref{fig:fig2}]. Respiratory complexes, among the machines driving the mitochondria out of equilibrium, operate on millisecond timescales. The required $P_a$ is lowered further by fluctuations with longer correlation times, thus entering the realm of biological plausibility. For comparison, heating a mitochondria-sized (1 \textmu m diameter) sphere in water to maintain a steady-state $10$ K temperature difference would require $\sim10^7$ pW, $10^5$ times the cellular energy budget.

\subsection*{Outlook}
While our model is too simple to quantitatively characterize any real mitochondrial temperature probe or the actual fluctuation spectrum of the mitochondrial interior, it nonetheless illustrates the physics of our proposed explanation for the hot mitochondria paradox. The cellular environment is home to nonequilibrium fluctuations, yielding a frequency-dependent effective temperature. Different temperature probes are sensitive to different parts of the fluctuation spectrum, such that even a perfectly calibrated probe can report seemingly impossible effective temperatures in nonequilibrium environments. The measured effective temperature difference between mitochondria and the rest of the cellular environment is thus both physically possible, and attainable with a biologically reasonable power budget.

We hope that future work will seek to test our theory experimentally. Our hypothesis is agnostic to the specific mechanism(s) responsible for generating nonequilibrium fluctuations inside the mitochondria. Indeed, mitochondria contain many active processes which could, and likely do, generate such fluctuations, including the rotation of ATP Synthase subunits on a millisecond timescale, and various nonequilibrium electrochemical reactions over a wide range of timescales. Future work should seek to model and experimentally characterize fluctuations within the mitochondria.

Equally important will be quantitatively modelling the probes used to measure mitochondrial temperature to determine both the response function $\chi(\omega)$ for their observed coordinates and the resulting probe kernel $g(\omega)$ given the measured observables (including nonlinear state-depedent fluorescence for fluorescent probes). Characterizing the quantities involved in \eqref{eq:Tmeasdef} will enable prediction of $T_\mathrm{meas}$ given the probe and environment. More qualitatively, our hypothesis predicts that probes with different relaxation timescales will produce different temperature readouts, and we expect slower-relaxing probes to measure larger effective temperatures than faster probes. A preliminary experiment could use several probes with a range of relaxation timescales to characterize effective temperature across frequencies.

We conclude by noting that nonequilibrium fluctuations in the cellular environment are not just a nuisance for thermometry; they also present an opportunity. Molecular machines with coupled components sensitive to different parts of the fluctuation spectrum would experience an effective temperature difference. This constitutes a thermodynamic resource that could be leveraged to perform useful work~\cite{leighton2024information,buisson2025hunting}.

\emph{Acknowledgements}.---We thank members of the Machta group for helpful discussions, Damon Clark for feedback on the manuscript, and Yale and NSERC for postdoctoral fellowship support.

\bibliography{main}

\end{document}